\documentclass[a4paper,10pt, twocolumn]{article}
\date{}
\usepackage[latin1]{inputenc}
\usepackage{amsmath}
\usepackage{amsfonts}
\usepackage{graphicx}
\usepackage{pgf}
\usepackage{amssymb}
\usepackage{textcomp}

\title{\textbf{Software-Hardware Evolution and birth of\\
 Multicore Processors}}
 
\author{K.R. Chowdhary, Professor\\
\small Dept. of Computer Science and Engineering,\\
\small JNV University, Jodhpur.\\
\small \textsf{Email:kr.chowdhary@acm.org}
}
\begin{document}

\maketitle

\begin{abstract}
\noindent
\textit{This paper presents a brief journey to the evolution of computer hardware and software, and underlines that shift to multicore technology is natual part of the evolution, and highlights the various laws governing the advancement of computer industry. Looking to these, it appears that the HW-SW industry trend can be represented by a mathematical model, for which future developments are predictable. Finally, the paper establishes that future of computer industry lies in more thrust in software to exploit parallelism available in software to utilize the heterogeneity in multicore processors.
}\\ 

\noindent
\textbf{Keywords:} Multicore, Moore's law, Amdahl's law, Myhrvold's law, SW-HW heterogeneity.  
\end{abstract}
 
\section{Introduction}
The computer performance has been driven by largely decreasing the size of the chips while increasing the number of transistors they contain. In accordance with Moore's law this has caused the chipmakers to rise and prices to drop. This ongoing trend has driven the computer industry for years\cite{dg}.

However, the transistors cannot shrink forever. As transistors components grow thinner, chip manufacturers find it difficult to cap power usage and heat generation. Resulting to this, manufacturers have started building chips with multiple cores, each separately cooled, instead of one increasingly powerful core. The multiple core chips do not necessarily run as fast as the highest performing single-core models, but they improve over all performance by handling more work in parallel.

The past progress was inefficient in terms of transistors and and power (such as multiple instruction issue, deep pipeline, out-of-order execution, speculative execution, and prefetching), but that increased performance while preserving the sequential programming model.

Multicore chips are biggest change in PC programming since 32-bit 386 architecture was introduced in 1990s. The multicore are a way to extend the Moore's law so that the user gets more performance out of a piece of silicon. The  multicore chips have been introduced by AMD, IBM, Intel and Sun micro in servers, desktops, and laptops. 

\subsection{Driving multicore}
Current transistor technology limits the ability to continue making single processor cores more powerful. For example, as transistor gets smaller, the gate which switches the electric current due to electrons - off and on, gets thinner and less able to block the flow of electrons (figure-1). Thus small transistors tend to use electricity all the time, even when they are not switching. This wastes and dissipates the power. Also increasing clock speeds causes transistors to switch faster and thus consume more power and generate more heat.

As per Moore's law, the clock rates would have exceeded 10 GHz in 2008, and 15 GHz in 2010, if there were no switch over to multicore by chip manufacturers in 2005. Due to this shifting to multicore multicore, the clock rate has been restrained below 3 GHz in 2010, which is less than that of single core processors in 2004, helping substentially in reduction of power dissipation and heating. 

\begin{figure}[t]
\begin{center}
 \includegraphics[width=5cm,height=3cm,bb=14 14 431 206,clip=true,trim=0pt 0pt 0pt 0pt]{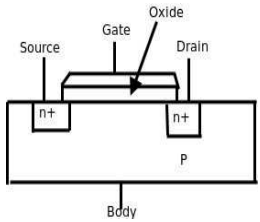}
 \caption{P-channel FET with gate, drain, and source connections}
\end{center}
\end{figure}

\subsection{Inside multicore}
Consider a dual core chip running multiple applications against a single core. Each core in multicore chip includes every thing a multiprocessor has, except level-2 cache memory hierarchy, which is outside the multicore.

In a muticore system, compiler handles the scheduling of instructions in a program, while operating system (OS) controls the over all assignment of tasks. The OS or multithread application parcels out the work to multicores. Generally, when a multicore processor has completed a task, one core takes the completed data  from the other cores and assembles the final result.

\subsection{Applications}
To take advantage of multicore chips, vendors must redesign applications. So that processor can run them as multiple threads. The programmers must find good places to breakup the applications, divide the work into roughly equal pieces that can run at the same time.

The vendors must redesign applications so that they can recognize each core's speed and memory access capabilities as well as how fast cores can communicate with one another.    

The typical multicore chips are:
AMD, IBM: (Power p5, p6, p7), Intel: Xeon, Sun: Niagara processor - 8 core (has shared II level cache).

The multicore processors will find a natural home in servers, but would not be very useful in desktops until vendors develop considerable more multithreaded applications\cite{jt}.

If Moore's law continues to apply, the number of cores in a chip will keep on increasing. The server applications primarily focus on throughput per cost and power. The multicore targeted for these applications use large number of small low power cores. However, desktop users are interested in the performance of a single application at a time. The multicore for desktop users are more likely to have smaller number of larger, higher power cores with better single thread performance. Thus, general solution is heterogeneous chip microprocessor with both high and low complexity cores\cite{rk}.  

\subsection{Advantages}
There is a speed gain when multiple tasks are run on multicore, compared to on the single-core CPU. Because cores are on the same die, they can share architectural components such as memory elements and memory management. They thus have fewer components and lower costs than systems running multiple chips. Also, signaling between cores can be faster and use less electricity than multichip systems. 

\section{Hardware Innovations and Moore's Law}
In 1965, Gorden Moore, a co-founder of Intel, postulated that number of transistors that could be fabricated on a semiconductor chip would double every year. Amazingly, this forecast still holds. Each next generation of transistor is smaller and switches at faster speed, allowing clock speed and computer performance to increase at a similar rate (figure 2, table 1). The figure shows that number of transistors increase at the rate predicted by Moore's law, but clock frequency grew at slightly a different rate~\cite{jl}.
 
\begin{table}
\caption{Performance and clock rate increase in Intel processors}
\begin{center}
\begin{tabular}{llrl}
\textbf{Year} & \textbf{Processor} & \textbf{Transistors} & \textbf{Clock f.}\\
1978 & 8086 & $29\times10^{3}$ & 5.00Mhz\\
2006 & Intel core2d & $291\times10^{6}$ & 2.93GHz
\end{tabular}
\end{center}
\end{table}

\begin{figure}[t]
\begin{center}
 \includegraphics[width=9cm,height=4cm,bb=14 14 431 206,clip=true,trim=0pt 0pt 0pt 0pt]{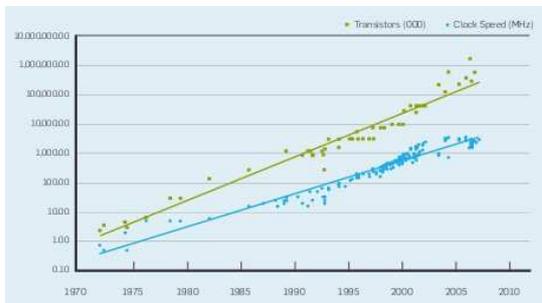}
 \caption{Improvement in Intel86 architecture over the years\cite{jl}.}
\end{center}
\end{figure}

Faster processors governed by Moore's law enabled software vendors to add new features and functionality to software that in turn demanded larger developer teams. The challenges of constructing increasingly complex software increased demand for high level languages and program libraries. Their higher level of abstraction contributed to slower code and, in conjunction with larger and more complex programs, drove demand for faster processors and closed the cycle, as indicated in figure 3. 

\begin{figure}[t]
\begin{center}
 \includegraphics[width=9cm,height=4cm,bb=14 14 431 206,clip=true,trim=0pt 0pt 0pt 0pt]{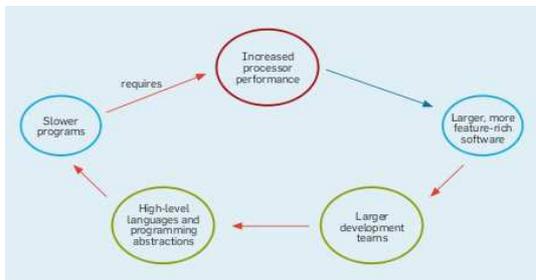}
 \caption{Cycle of HW-SW Innovation process in Computer Industry\cite{jl}.}
\end{center}
\end{figure}

Now the era of steady growth of single processor performance is over, and there is a transition from sequential to parallel computation (multicore). The sequential computing era ended when the practical limits on power dissipation stopped the continual increase in clock speed, and lack of exploitation in instruction level parallelism diminished the values of complex architectures.

However, the Moore's law is still applicable, as semiconductor technology still doubles the number of transistors on chip every two years\cite{hm}. This doubling of transistors is now used to increase the number of independent processors on chip, instead of attempting to enhance the capability of individual processor. Now the challenge is - how to design software to exhibit the capability of multicore architecture. In other words, will parallelism continue the cycle of software innovation?

\section{Software Innovations and Myhrvold's Law}
A common belief among software developers is that software grows at least at the same rate as the platform on which it runs. Nathan Myhrvolds, former chief technology officer at Microsoft, memorably captured this wisdom with his four laws about software\cite{nm}.

\begin{enumerate}
 \item \emph{Software is like a gas, which has tendency to expand to fill the capacity of any computer}. 
\begin{itemize}
\item Windows NT: LOC (lines of code) doubling time 866 days (@ 33.90 percent per year).
\item Browser: LOC doubling time 216 days, growth rate 221 percent per year.
\end{itemize}

\item \emph{Software grows until it comes to limitation by Moore's law.}\\
The initial growth is quick, like gas expanding (as in browser), eventually limited by hardware (as in NT), which ultimately brings any processor to its knees, just before the new model is out.

\item \emph{Software growth makes Moore's law possible.}\\
That is why chips get faster at the same price, not cheaper. This will continue as long as there is opportunity for new software.

\item \emph{Impossible to have enough new algorithms, new applications, and new notion of what is cool.}\\
\end{enumerate}

In fact, increase in size of SW has not only overloaded the processor and memory, but have many positive effects also. Some of these are following:\\
 
\begin{enumerate}
 \item \emph{Increased functionality}
 \begin{itemize}
\item Improved security
\item Printer and IO drivers for graphics
\item Printer resolution and color depth (24 bits)
\item Improved software engineering practices such as layering of SW architecture and modularizing system to improve development
\item The data manipulated by computers also evolved from simple ASCII to larger structured objects (word and excel documents), to compressed documents like jpeg, xml (computation and space saving)
\item Growing use of videos
\end{itemize}
 
\item \emph{Programming changes}\\
Over the last 30 years, programming languages have evolved from Assembly language and C code to high level languages. Major step was C++, which brought object oriented mechanism. C++ also introduced abstraction, like classes and template, and made possible rich libraries. These features give expensive run time implementations, modularity in development, and information hiding. These practices enabled the creation of ever larger and more complex software.

\hspace{2mm} Safe and managed languages such as $C\sharp$ and Java further increased the level of programming by introducing garbage collection, richer class libraries (.net, java classes), just-in-time compilation, etc. All these features provide powerful abstractions for developing software but consume lot of processor resources and memory.

\hspace{2mm}The rich features of language requires a run time system to maintain a large amount of meta-data on every method and class at run time, even if the required features are not invoked. The table 2 shows the code for compiling \textquotedblleft Hello World\textquotedblright in Windows Vista, coded using Visual studio. The table 3 shows the execution time of \textquotedblleft Hello World\textquotedblright.

\begin{table}
\caption{Compiled Code size for \emph{Hello World}}
\begin{center}
\begin{tabular}[c]{lr}
Language & Size in bytes\\
\hline
C & 5874\\
C++ & 8762
\end{tabular}
\end{center}
\end{table}

\begin{table}
\caption{Execution time for \emph{Hello World}}
\begin{center}
\begin{tabular}[c]{lr}
Mechanism & Timer(280 nsec.)\\
\hline
C++, Console & 1760\\
C++, Windows & 36375\\
$C\sharp$, console & 2628\\
$C\sharp$, windows & 80348
\end{tabular}
\end{center}
\end{table}

\item \emph{Decreased Programming Focus}\\
Abundant machine resources have allowed programmers to become easy-going about performance and less aware about resource consumption by their code. More important is change in the mindset of the developer. Consequently, most codes run at near machine capacity, and any further increase in code size / complexity will retard the performance.

\end{enumerate}
 
\section{Mitigating Amdahls's Law}
The Amdahl's law sates that speed of a particular application is limited by the fraction of the application that is serial (cannot be parallelized). During the serial portion of execution, the chip's power budget is applied towards using a single large core to allow the serial portion to execute as quickly as possible. During the parallel portions, the chip's power budget is used more efficiently by running the parallel portions on  large number of small-area and power efficient cores. Thus, executing serial portions of an application on a fast but relatively inefficient core and executing parallel portions of an algorithm on many small cores can maximize the ratio of performance to power dissipation.\\

\noindent
\textbf{Illustrative Example:}
Let for 10 percent of time a program gets no speedup on a 100-core computer. To run this sequential piece twice as fast, assume a single fat core would need 10 times as many resources as a thin core due to larger caches, a vector unit, and other features. Applying Amdahl's law, the speedups can be computed as follows\cite{ka}:

\begin{itemize}
 \item[i] True serial: 1 sec.
 \item[ii] For 100 thin cores:\\
  Required time = 0.1 + 0.9/100 = 0.109 sec.\\
  Relative speedup = 1/0.109 = 9.17 times
 \item[iii] For 90 thin cores, and one thick core:\\
  Required time = 0.1/2 + 0.9/90 = 0.06 sec.\\
  Relative speedup = 1/0.06 = 16.66 times
\end{itemize}

Since, the implementation cost of one thick core is taken as equal to 10 thin cores, the total cost in (ii) and (iii) is going to be same.\\

\noindent
\textbf{Requirement of Heterogeneity in Software:}
To take full advantage of heterogeneous Chip Multiprocessors (CMPs), the system software must use the execution characteristics of each application to predict its future processing needs and then schedule it to a core that matches those needs if one is available. The predictions can minimize the loss to the whole system rather than to the single application. 

Effective schedulers can be implemented, even with the current commercial operating systems, and open OS like Linux.

To achieve the best performance it will require to compile the program for heterogeneous CMPs slightly differently. That is, for statically schedulable and dynamically schedulable cores. Programming or compiling parallel applications might require more awareness of heterogeneity. Application developers assume that cores provide equal performance, but heterogeneity breaks this assumption. In fact, many of the future areas of research remain for heterogeneous CMPs\cite{sb}. 

\section{Multicore and its future}

With clock speeds stalling out, and computational power increasing due to doubling of number of cores per processor, serial computing is now dead, and parallel computing revolution is now upon us. However, writing paralle applications is a significant challenge. For parallelism to succeed it must ultimately produce better performance relative to speed, efficiency, and reliability. However, most programmars are not only ill equipped to produce proper parallel programs, they also lack the tools, and environments for producing such programs\cite{wf}. 

Dealing with those issues require a suite of tools and environments that provide users and developers with convenient mechanisms for managing different resources in multicore environments. The resources are memory, cache, and computing units, compilers that allow sequential programs to automatically take advantages of multicore systems, strategies that allow users to analyse performance issues in multicore systems, and environment to control and eliminate bugs in parallel threaded programs.

Developing parallel algorithms are considerable challenge, but many problems, such as video processing, natural language interactions, speech recognition, linear and nonlinear optimization, machine learning, are some of the areas, which are computationally intensive, and multicore are the solutions for these. However, Amdahl's law limit the scope until sequential component itself is rescheduled.

An alternative use for multicore processor is to redesign a sequential application into a loosely coupled or asynchronous system in which computations run on separate processors. For example, it is natural to separate monitoring features from program logic.

It is fact that, for many applications, most functionality is likely to remain sequential. For software developers to find the resources to add or change features, it may be necessary to eliminate old features or reduce their resource consumption. An important consequence of multicore is that sequential performance tuning and code-restructuring tools are likely to be increasingly important.

Another challenge is modifications and updating of parallel code. Parallelism will also force major change in software development. Moore's dividend enabled a shift to higher level languages and libraries, the pressure driving this trend will not change, because increasing abstraction improves security reliability, and program productivity. In multicore, parallel garbage collection does not increase pause time of threads, hence real-time applications are easier to implement with real-time constraints.

Another approach that sacrifices the performance for development productivity is, to hide the underlying parallel implementation.

\section{Tools for Multicore programming}
Cetus \cite{cd,pd} is an open source source-to-source C compiler written in java and maintained at Purdue University. The Cetus has features of automatic parallelization of C code. The following part shows the transformation of C code by Cetus.

\begin{verbatim}
#Input to Cetus
int temp;
int main(void) {
   int i,j, c, a[100];
   c = 2;
   for (i=0; i<100; i++){
      a[i] = c*a[i] + (a[i] -1);
  }
} 

#Output from Cetus
int temp;
int main(void)
 { 
   int i,j, c, a[100];
   c = 2;
   for (i=0; i<100; i++)
     {
      a[i] = ((c*a[i]) + (a[i] -1));
  }
} 
\end{verbatim}

Cetus performs symbolic transformations for efficient source code generation and the one which can be parallelized easily. Some examples are:\\

  1 + 2*a +4 -a  $\Rightarrow$ 5 + a (folding)\\
  a*(b + c) $\Rightarrow$ a*b + a*c  (distribution)\\
  (a*2)/(8*c) $\Rightarrow$ a / (4*c) (division)\\

\section{Multicore programming languages}  
A parallel programming language must have key properties. First, expressing parallelism should be simple. Second, should be able to combine number of parallel programming models. Third, the languages must exploit parallel resources efficiently. \\
Cilk, an extension of C provides two key words for expressing parallelism. A \textit{spawn} transforms sequential (blocking) function call into an asynchronous (non blocking) call. A \textit{sync} blocks a function's execution until all its spawned children are completed. 

\begin{verbatim}
 Cilk int fib (n) {
    if ( n < 2 ) return n;
    else
       int x, y;
       x = spawn fib (n -1);
       y = spawn fib (n -2);
       sync;
        return ( x + y);
  }
}
\end{verbatim}

Because the recursive calls to fib are spawned, Cilk's runtime system can execute them in parallel. Because the expression (x + y) depends on the results of both calls, the sync ensures that both have completed before the addition begins.

The Cilk's run time system must efficiently map logically independent calls onto computational cores.
 
\section{Conclusion}
The multicore processors will change the software profoundly as previous hardware evolutions such as the shift from vacuum tubes to transistors or transistors to ICs. Parallelism will drive software in new direction which is computational intensive.

\end{document}